\documentclass[preprint,showpacs,preprintnumbers,amsmath,amssymb]{revtex4}

\usepackage{graphicx}
\usepackage{dcolumn}
\usepackage{bm}

\begin{document}

\title{Theory of the nonsteady diffusion growth of a gas bubble\\
in a supersaturated solution of gas in liquid}

\author{A. P. Grinin}
\author{F. M. Kuni}
\author{G. Yu. Gor}
 \email{gennady_gor@mail.ru}
\affiliation{
Saint-Petersburg state university, Research institute of physics \\
198504 Russia, St. Petersburg, Petrodvorets, Ulyanovskaya st., 1
}%

\date{\today}

\begin{abstract}
Using a self-similar approach a general nonsteady theory is elaborated for the case of the diffusion growth of a gas bubble in a supersaturated solution of gas in liquid. Due to the fact that the solution and the bubble in it are physically isolated, the self-similar approach accounts for the balance of the number of gas molecules in the solution and in the bubble that expells incompressible liquid solvent while growing. The rate of growth of the bubble radius in its dependence from gas solubility and solution supersaturation is obtained. There is a nonsteady effect of rapid increase of the rate of bubble growth simultaneous with the growth of the product of gas solubility and solution supersaturation. This product is supplied with a limitation from above, which also stipulates isothermal conditions of bubble growth. The smallness of gas solubility is not presupposed.
\end{abstract}


\maketitle

\section*{Introduction}
\label{Introduction}

The basis of the study of the diffusion growth of a gas bubble in a supersaturated solution of gas in liquid is usually \cite{Kuni_Ogenko_1, Trofimov_Melikhov_Kuni, Kuni_Zhuvikina_2002, Kuni_Zhuvikina_Grinin, Slezov_1, Slezov_2} founded on a steady solution of the diffusion equation of gas molecules in liquid. It is the purpose of this paper to elaborate a general nonsteady theory for the diffusion growth of a gas bubble in a supersaturated solution of gas in liquid, exploiting a self-similar approach elaborated in \cite{Vasilyev} by the example of the diffusion growth of a liquid droplet in a supersaturated vapor-gas mixture. This theory does not exploit the supposition that the concentration profile of gas dissolved around the growing bubble is steady.

The object under consideration is a liquid solution of gas which includes a gas bubble of the same single-component substance that is dissolved in the liquid. We shall consider the volume of the solution to be so large that heterogeneous nucleation of bubbles at its boundaries does not influence the volume of the solution where the bubble is located. The bubble is rather large, which makes it possible to neglect the Laplace forces. The solution is presupposed to be diluted, while the gas in the bubble -- perfect. Dissociation and chemical transformations of the dissolved molecules are also neglected. Due to high heat conductivity of the liquid solution, we shall consider all the system of the solution and of the bubble as having the same temperature. Temperature and pressure in the solution are given. The solubility of gas in liquid is traditionally seen as a dimensionless value defined (at given temperature and pressure) as the ratio between the volume of gas and the volume of liquid dissolved in this gas.

We shall study a realistic situation when the diffusion of molecules of the dissolved gas in the presence of a bubble occurs simultaneously with the motion of the liquid solvent which is caused by the movement of the bubble surface and due to the solvent incompressibility. This problem has not been studied in \cite{Kuni_Ogenko_1, Trofimov_Melikhov_Kuni, Kuni_Zhuvikina_2002, Kuni_Zhuvikina_Grinin, Slezov_1, Slezov_2}. Taking into account the fact that the solution and the bubble in it are physically isolated, we shall consider the balance of the number of gas molecules in the solution and in the bubble that expells the incompressible liquid solvent while growing. Using the self-similar theory as a basis, we shall derive an equation for the rate of growth of the bubble radius in its dependence on gas solubility and solution supersaturation. We shall describe a significantly nonsteady effect of a rapid increase of the rate of bubble growth at the increase of the product of gas solubility and solution supersaturation. This product will be provided with the limitation from above, which stipulates the isothermal conditions of bubble growth. The smallness of gas solubility will not be presupposed.

\section{Gas bubble growing in liquid solution of gas: general ideas}
\label{GeneralIdeas}

The state of the solution is stipulated by temperature $T$, pressure $\Pi $ and the initial concentration (the number density of molecules) $n_{0} $ of the dissolved gas. The meaning of value $n_{0} $ will be revealed further in the initial condition in Eq. \eqref{EQ_1_12_} and in the boundary condition in Eq. \eqref{EQ_1_15_}. $n_{\infty } $ will stand for the concentration of the dissolved gas in the saturated solution, which at given temperature $T$ and pressure $\Pi $ is characterized by chemical and mechanical equilibrium with the clean gas above the plane contact surface. We shall consider the solution to be diluted. The solution contains a gas bubble (the same gas that is dissolved in the solution). Due to high heat conductivity of the liquid solvent, the bubble of gas will have the same temperature $T$ as the solution does. The condition that the latter is valid and that the temperature in the bubble is homothermal will be justified in section \ref{Applicability}. There we shall also justify the mechanical equilibrium of the bubble and the solution. The radius of the bubble will be designated as $R$.

We shall suppose the radius $R$ to be that large that the following strong inequalities are observed: 
\begin{equation} \label{EQ_1_1_} 
R{\rm \gg }2\sigma /\Pi , 
\end{equation} 
\begin{equation} \label{EQ_1_2_} 
R{\rm \gg }4sD/\alpha _{c} {\rm v}_{T} . 
\end{equation} 
Here $\sigma $ is the surface tension of the liquid solvent (in the case of a diluted solution), $D$ is the diffusion coefficient of gas molecules in the liquid solvent, $\alpha _{c} $ is the condensation coefficient of a gas molecule at the virtual transition of gas molecules from the bubble and into the solution \cite{Kuni_Ogenko_1}, ${\rm v}_{T} $ is a mean heat velocity of a gas molecule inside the bubble, $s$ is the gas solubility that is defined as a dimensionless value via
\begin{equation} \label{EQ_1_3_} 
s\equiv kTn_{\infty } /\Pi , 
\end{equation} 
where $k$ is the Boltzmann constant.

When Eq. \eqref{EQ_1_1_} is observed, the influence of the Laplace forces on the bubble is rather small and the pressure of gas inside the bubble at its mechanical equilibrium with the solution is equal to pressure $\Pi $ of the solution. At the same time the gas inside the bubble can be considered perfect; therefore for gas concentration in the bubble, that is designated as $n_{g} $, we have
\begin{equation} \label{EQ_1_4_} 
n_{g} =\Pi /kT, 
\end{equation} 
so $n_{g} $ is unambiguously defined by temperature $T$ and pressure $\Pi $ of the solution. Using Eq. \eqref{EQ_1_3_}, we can also rewrite Eq. \eqref{EQ_1_4_} as 
\begin{equation} \label{EQ_1_5_} 
n_{g} =n_{\infty } /s. 
\end{equation} 
Due to Eq. \eqref{EQ_1_5_} value $s$ defined in Eq. \eqref{EQ_1_3_} corresponds to the usual understanding of solubility, that is, a ratio (at given temperature and pressure) of the volume of gas to the volume of liquid that has dissolved this gas. Solubility $s$ and value $n_{\infty } $ depend not only on temperature $T$ and pressure $\Pi $, but also on the type of solvent and on the type of dissolved gas.

We shall neglect the fugacity of the liquid solvent and do so for the following reason: let $P_{\beta } $  be the pressure of the saturated vapor of the solvent. If we have a look at the thermodynamic diagram, at temperature $T$and pressure $\Pi $ of the solution the solvent will be found deep within the area of the steadiness of the liquid phase. For this reason $P_{\beta } {\rm \ll }\Pi $, which, using concentration $n_{\beta } $ of the saturated vapor of the liquid solvent, can be rewritten as $n_{\beta } {\rm \ll }\Pi /kT$. From this and from Eq. \eqref{EQ_1_4_} it follows that $n_{\beta } {\rm \ll }n_{g} $. Thus the bubble consists mainly of gas; that is why the fugacity of the solvent can in fact be neglected.

As it was shown in \cite{Kuni_Zhuvikina_2002}, the strong inequality in Eq. \eqref{EQ_1_2_} is the condition for the existence of the diffusion regime of bubble growth. Leaving a certain reserve, let us assume that
\begin{equation} \label{EQ_1_6_} 
R_{0} {\rm \sim }20\max \left\{\frac{2\sigma }{\Pi } ,\frac{4sD}{\alpha _{c} {\rm v}_{T} } \right\}. 
\end{equation} 
Eq. \eqref{EQ_1_6_} evaluates value $R_{0} $ of the bubble radius, which is initial for the existence of conditions in both Eq. \eqref{EQ_1_1_} and Eq. \eqref{EQ_1_2_}. Jointly with those, Eq. \eqref{EQ_1_4_} and the diffusion regime of the bubble growth are also well observed. The largest of the two values given in braces in Eq. \eqref{EQ_1_6_} is, as a rule, the value of $2\sigma /\Pi $.

Let us introduce characteristic time $t_{0} $, at which the following initial condition on the radius of the bubble is observed: 
\begin{equation} \label{EQ_1_7_} 
R(t)|_{t=t_{0} } =R_{0} . 
\end{equation} 
Time $t$ is calculated from the time of nucleation of the bubble that is growing irreversibly. From Eq. \eqref{EQ_1_7_} and the meaning of value $R_{0} $ it can be seen that after time $t_{0} $ after the nucleation of the bubble Eq. \eqref{EQ_1_4_} and the diffusion regime of bubble growth will be observed simultaneously. The larger $R_{0} $, that is, the more precise is the observation of Eqs. \eqref{EQ_1_1_} and \eqref{EQ_1_2_}, the larger $t_{0} $.

The bubble perturbs the surrounding solution. Let us designate the concentration of the dissolved gas at distance $r$ from the centre of the bubble at time $t$ as $n(r,t)$. If the growth regime of the bubble is a diffusion one, the evolution equation is observed
\begin{equation} \label{EQ_1_8_} 
\frac{\partial n(r,t)}{\partial t} =D\Delta n(r,t)-{\rm div}\left[n(r,t)\vec{{\rm v}}(\vec{r},t)\right]\; \; \; \; \; (t{\rm \geqslant }t_{0} ). 
\end{equation} 
The first member in Eq. \eqref{EQ_1_8_} accounts for the diffusion of gas molecules, while the second member stipulates it that this diffusion occurs against the background of the movement of liquid solvent caused by the movement of the surface of the bubble and by the incompressibility of the solvent. If we study the velocity of solvent movement designated in Eq. \eqref{EQ_1_8_} as $\vec{{\rm v}}(\vec{r},t)$ in the spherical system of co-ordinates with the reference point in the centre of the bubble, we shall see that it has only radial component ${\rm v}_{r} (r,t)$, for which, as it is obvious, the following is valid: 
\begin{equation} \label{EQ_1_9_} 
{\rm v}_{r} (r,t)=\frac{R^{2} (t)}{r^{2} } \frac{dR(t)}{dt} . 
\end{equation} 
Using Eq.\eqref{EQ_1_9_} and some well-known formulas of vector operations
\begin{equation} \label{EQ_1_10_} 
\Delta f\equiv \frac{1}{r^{2} } \frac{\partial }{\partial r} \left(r^{2} \frac{\partial f}{\partial r} \right),\; \; {\rm div}\vec{A}\equiv \frac{1}{r^{2} } \frac{\partial }{\partial r} \left(r^{2} A_{r} \right) 
\end{equation} 
let us rewrite the evolution equation from Eq. \eqref{EQ_1_8_} as follows
\begin{equation} \label{EQ_1_11_} 
\frac{\partial n(r,t)}{\partial t} =\frac{D}{r^{2} } \frac{\partial }{\partial r} \left[r^{2} \frac{\partial n(r,t)}{\partial r} \right]-\frac{R^{2} (t)}{r^{2} } \frac{dR(t)}{dt} \frac{\partial n(r,t)}{\partial r} \; \; \; \; \; (t{\rm \geqslant }t_{0} ). 
\end{equation} 

At all $r>0$ we have the following initial condition on the homogeneity of the concentration profile of the dissolved gas at time $t=0$ of the bubble nucleation: 
\begin{equation} \label{EQ_1_12_} 
n(r,t)|_{t=0} =n_{0} \; \; \; \; \; (r>0). 
\end{equation} 
This condition brings about value $n_{0} $ as the initial concentration of the dissolved gas. In a supersaturated solution it is observed that $n_{0} >n_{\infty } $.

During the diffusion regime of the bubble growth we have an equilibrium boundary condition on the surface of the bubble:
\begin{equation} \label{EQ_1_13_} 
n(r,t)|_{r=R(t)} =n_{\infty } \; \; \; \; \; (t{\rm \geqslant }t_{0} ). 
\end{equation} 
We shall neglect the dependence of the right side of Eq. \eqref{EQ_1_13_} on the radius of the bubble when this radius is significant and conforms with the limitation from Eq. \eqref{EQ_1_1_}.

Let us impose a boundary condition of the absence of an external source of the dissolved gas: 
\begin{equation} \label{EQ_1_14_} 
r^{2} \left. \frac{\partial n(r,t)}{\partial r} \right|_{r=\infty } =0. 
\end{equation} 
This condition means that the solution and the bubble in it are physically isolated.

In accordance with Eqs. \eqref{EQ_1_12_} and \eqref{EQ_1_14_} it is also true that 
\begin{equation} \label{EQ_1_15_} 
n(r,t)|_{r=\infty } =n_{0} . 
\end{equation} 
Due to physical isolation of the solution and the bubble in it, value $n_{0} $ also has the meaning of concentration of the dissolved gas at an infinite distance from the bubble; and this value is constant in time.

Each element of the surface of the bubble of radius $R$ is moving with the velocity of the movement of the incompressible liquid solvent. Thus, 
\begin{equation} \label{EQ_1_16_} 
n_{g} \frac{dR(t)}{dt} =D\left. \frac{\partial n(r,t)}{\partial r} \right|_{r=R(t)} \; \; \; \; \; \; (t{\rm \geqslant }t_{0} ). 
\end{equation} 
Eq. \eqref{EQ_1_16_} can be interpreted simply. It corresponds to the diffusion flux of the numbers of molecules through the surface of radius $R$ (there is no convection flux due to mobility of this surface). The diffusion flux is spent on the change in the number of molecules of gas in the bubble (which at $t{\rm \geqslant }t_{0} $ , in concordance with the initial condition from Eq. \eqref{EQ_1_7_}, has constant concentration $n_{g} $ of pure gas). As the sign of the derivative by $r$ in Eq. \eqref{EQ_1_16_} obviously corresponds with the sign of value $n_{0} -n_{\infty } $, then, in the case of supersaturated solution, Eq. \eqref{EQ_1_16_} leads to
\begin{equation} \label{EQ_1_17_} 
dR/dt>0\; \; \; \; \; \; (t{\rm \geqslant }t_{0} ) 
\end{equation} 
-- i. e. leads to the growth of the bubble with time.

In order to specify the state of the solution one can use value $\zeta $ instead of value $n_{0} $. Value $\zeta $ can be defined as 
\begin{equation} \label{EQ_1_18_} 
\zeta \equiv (n_{0} -n_{\infty } )/n_{\infty } . 
\end{equation} 
In a supersaturated solution, where $\zeta >0$, value $\zeta $ will be referred to as the supersaturation of a solution.

\section{Self-similarity of the theory}
\label{SelfSimilarity}

Following \cite{Vasilyev}, let us introduce a self-similar variable $\rho $ having assumed that 
\begin{equation} \label{EQ_2_1_} 
\rho =\frac{r}{R(t)} \; \; \; \; \; (\rho {\rm \geqslant }1). 
\end{equation} 
We shall look for the solution $n(r,t)$ of the evolution equation \eqref{EQ_1_11_} in the form of $n(\rho )$ of one variable $\rho $: 
\begin{equation} \label{EQ_2_2_} 
n(r,t)=n(\rho ). 
\end{equation} 
Here and further, for the sake of simplicity, we shall not mention that $\rho $ is changing within the limits of $\rho {\rm \geqslant }1$.

In accordance with Eqs. \eqref{EQ_2_1_} and \eqref{EQ_2_2_} we have 
\begin{equation} \label{EQ_2_3_} 
\frac{\partial n(r,t)}{\partial t} =-\frac{r}{R^{2} (t)} \frac{dR(t)}{dt} \frac{dn(\rho )}{d\rho } , 
\end{equation} 
\begin{equation} \label{EQ_2_4_} 
\frac{\partial n(r,t)}{\partial r} =\frac{1}{R(t)} \frac{dn(\rho )}{d\rho } . 
\end{equation} 
From Eqs. \eqref{EQ_1_16_} and \eqref{EQ_2_4_} we have 
\begin{equation} \label{EQ_2_5_} 
\frac{dR(t)}{dt} =\frac{D}{R(t)n_{g} } \left. \frac{dn(\rho )}{d\rho } \right|_{\rho =1} .\; \; \; \; \; (t{\rm \geqslant }t_{0} ). 
\end{equation} 
Let us rewrite Eq. \eqref{EQ_2_5_} as
\begin{equation} \label{EQ_2_6_} 
\frac{dR(t)}{dt} =\frac{Db}{R(t)} \; \; \; \; \; (t{\rm \geqslant }t_{0} ) 
\end{equation} 
or, equivalently, as
\begin{equation} \label{EQ_2_7_} 
dR^{2} /dt=2Db\; \; \; \; \; (t{\rm \geqslant }t_{0} ), 
\end{equation} 
where we introduce an important dimensionless parameter $b$ using 
\begin{equation} \label{EQ_2_8_} 
b\equiv \frac{1}{n_{g} } \left. \frac{dn(\rho )}{d\rho } \right|_{\rho =1} . 
\end{equation} 
In accordance with Eq. \eqref{EQ_2_7_}, the derivative $dR^{2} /dt$ does not depend on time: along the axis of variable $R^{2} $ the bubble is ``moving'' with the velocity that does not depend on time and on the size of the bubble.

Relative measure of inaccuracy of the self-similar velocity of the movement of the bubble $2Db$ in Eq. \eqref{EQ_2_7_} is equal to the relative measure of inaccuracy of Eq. \eqref{EQ_1_4_}, which, in its turn, does not exceed value $2\sigma /R_{0} \Pi $, which can be easily obtained from the estimate from Eq. \eqref{EQ_1_6_}.

It is easy to make sure that, if we take Eqs. \eqref{EQ_2_1_} -- \eqref{EQ_2_4_} and Eq. \eqref{EQ_2_6_} into consideration, the evolution Eq. \eqref{EQ_1_11_} turns into an ordinary differential equation for function $n(\rho )$ of the following type: 
\begin{equation} \label{EQ_2_9_} 
\frac{d^{2} n(\rho )}{d\rho ^{2} } +\left[\frac{2}{\rho } +b\left(\rho -\frac{1}{\rho ^{2} } \right)\right]\frac{dn(\rho )}{d\rho } =0. 
\end{equation} 
Integrating Eq. \eqref{EQ_2_9_}, obtaining a constant from the first integration using Eq. \eqref{EQ_2_8_}, and a constant from the second integration using the boundary condition from Eq. \eqref{EQ_1_13_}, formulated by virtue of Eqs. \eqref{EQ_2_1_} and \eqref{EQ_2_2_} as $n(\rho )|_{\rho =1} =n_{\infty } $, we have 
\begin{equation} \label{EQ_2_10_} 
n(\rho )=n_{\infty } +n_{g} be^{3b/2} \int _{1}^{\rho } \frac{dx}{x^{2} } {\rm e}^{-{bx^{2} \mathord{\left/ {\vphantom {bx^{2}  2-{b\mathord{\left/ {\vphantom {b x}} \right. \kern-\nulldelimiterspace} x} }} \right. \kern-\nulldelimiterspace} 2-{b\mathord{\left/ {\vphantom {b x}} \right. \kern-\nulldelimiterspace} x} } } . 
\end{equation} 
The fact that Eqs. \eqref{EQ_2_9_} and \eqref{EQ_2_10_} refer only to times $t{\rm \geqslant }t_{0} $ is not stipulated for the sake of simplicity. As it is obvious from Eqs. \eqref{EQ_2_1_} and \eqref{EQ_2_2_}, a self-similar formula \eqref{EQ_2_10_} complies with the boundary condition from Eq. \eqref{EQ_1_14_}, that is the absence of an external source of the dissolved gas.

In order to find parameter $b$ using Eq. \eqref{EQ_2_10_} and the boundary condition from Eq. \eqref{EQ_1_15_} formulated by virtue of Eqs. \eqref{EQ_2_1_} and \eqref{EQ_2_2_} as $n(\rho )|_{\rho =\infty } =n_{0} $, we come up with the following transcendental equation: 
\begin{equation} \label{EQ_2_11_} 
a=be^{3b/2} \int _{1}^{\infty } \frac{dx}{x^{2} } {\rm e}^{-{bx^{2} \mathord{\left/ {\vphantom {bx^{2}  2-{b\mathord{\left/ {\vphantom {b x}} \right. \kern-\nulldelimiterspace} x} }} \right. \kern-\nulldelimiterspace} 2-{b\mathord{\left/ {\vphantom {b x}} \right. \kern-\nulldelimiterspace} x} } } , 
\end{equation} 
where an important dimensionless parameter $a$ is introduced using 
\begin{equation} \label{EQ_2_12_} 
a\equiv \frac{n_{0} -n_{\infty } }{n_{g} } . 
\end{equation} 
In the case of a supersaturated solution, from Eq. \eqref{EQ_2_12_} it follows that
\begin{equation} \label{EQ_2_13_} 
a>0. 
\end{equation} 
In accordance with Eqs. \eqref{EQ_1_5_} and \eqref{EQ_1_18_}, we can rewrite Eq. \eqref{EQ_2_12_} as:
\begin{equation} \label{EQ_2_14_} 
a=s\zeta . 
\end{equation} 

In order that the integral in Eq. \eqref{EQ_2_11_} should converge and, simultaneously, that the right part of Eq. \eqref{EQ_2_11_} should be positive in concordance with Eq. \eqref{EQ_2_13_}, it is required that 
\begin{equation} \label{EQ_2_15_} 
b>0. 
\end{equation} 
Eqs. \eqref{EQ_2_6_} and \eqref{EQ_2_15_} have the previous inequality \eqref{EQ_1_17_} as a consequence.

Eq. \eqref{EQ_2_11_} for $b$ relative to $a$ has an unambiguous solution within the area stipulated by Eq. \eqref{EQ_2_13_}. The results of numerical solution of Eq. \eqref{EQ_2_11_} are represented by the curve in Fig. 1. This curve does not depend on the type of the solvent and on the type of the dissolved gas. All initial parameters of the problem are represented on the curve solely by means of parameter $a$ introduced in Eq. \eqref{EQ_2_12_}.

\begin{figure}
\includegraphics[width=180mm]{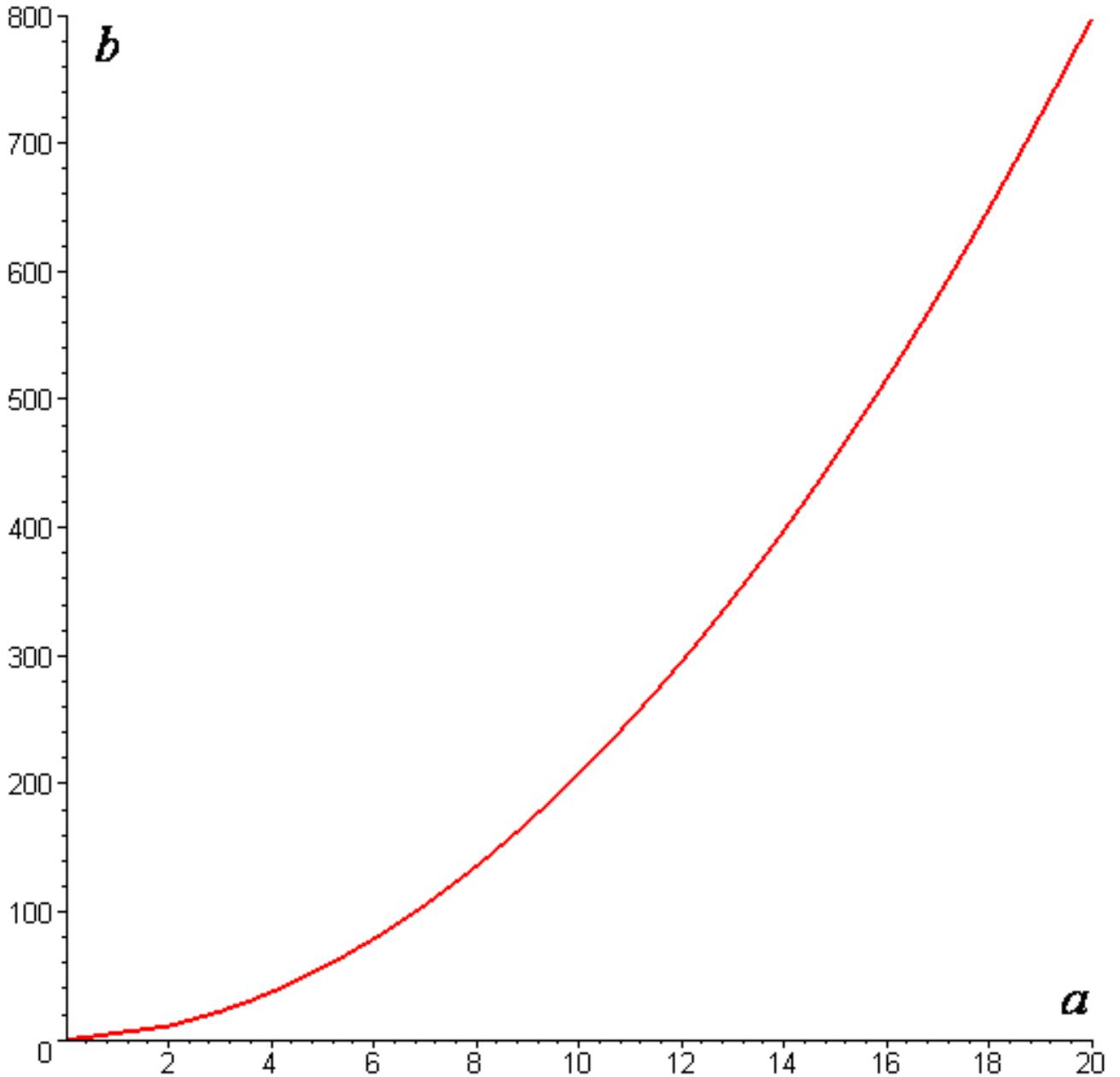}
\caption{Dependence of the solution $b$ of Eq. \eqref{EQ_2_11_} on parameter $a$. }
\label{Fig1}
\end{figure}

Integrating Eq. \eqref{EQ_2_7_} with the initial condition \eqref{EQ_1_7_}, we have 
\begin{equation} \label{EQ_2_16_} 
R^{2} (t)=2Db\cdot (t-t_{0} )+R_{0}^{2} \; \; \; \; \; (t{\rm \geqslant }t_{0} ). 
\end{equation} 
When $R_{0} $ is significant, which is required by the estimate from Eq. \eqref{EQ_1_6_}, the relative measure of inaccuracy $2\sigma /R\Pi $ of the self-similar rate of bubble growth $2Db$ in Eq. \eqref{EQ_2_7_} does not exceed $10\% $ at times $t>t_{0} /4$ (and does not exceed $5\% $ at times $t>t_{0} $). During the longest stage of the whole time interval $0<t<t_{0} $, that is $t_{0} /4<t<t_{0} $, the bubble, consequently, is growing practically with a self-similar rate $2Db$. For this reason we can use an approximate formula 
\begin{equation} \label{EQ_2_17_} 
t_{0} {\rm \simeq }R_{0}^{2} /2Db, 
\end{equation} 
where it is stipulated that the bubble is growing with a self-similar rate $2Db$ throughout the whole interval of $0<t<t_{0} $. From Eqs. \eqref{EQ_2_16_} and \eqref{EQ_2_17_} we have
\begin{equation} \label{EQ_2_18_} 
R^{2} (t){\rm \simeq }2Db\; t\; \; \; \; \; (t{\rm \geqslant }t_{0} ), 
\end{equation} 
which makes it possible to find$R(t)$ in Eq. \eqref{EQ_2_1_} at all times $t{\rm \geqslant }t_{0} $. The larger $R_{0} $, the more accurate Eqs. \eqref{EQ_2_17_} and \eqref{EQ_2_18_} are.

Let us now explicitly formulate the sequence in which the values that are found during an experiment are introduced within the theory. The state of the solution is given by its temperature $T$, pressure $\Pi $ and the initial concentration $n_{0} $ of the dissolved gas (introduced into the theory by means of the initial condition in Eq. (1.12)). Using the given $T$ and $\Pi $ one can obtain from an experiment (at a stipulated type of solvent and type of a dissolved gas) value $n_{\infty } $, and, knowing $T$, $\Pi $ and $n_{\infty } $, using definition from Eq. \eqref{EQ_1_3_}, we introduce the theory with gas solubility $s$. Then, using either of Eqs. \eqref{EQ_1_4_} or \eqref{EQ_1_5_}, we obtain value $n_{g} $ (that is unambiguously defined by temperature $T$ and pressure $\Pi $). Finally, knowing $n_{0} $, $n_{\infty } $ and $n_{g} $, by means of definition from Eq. \eqref{EQ_2_12_} we introduce value $a$.

It should be mentioned that in order to stipulate value $n_{0} $ in the experiment it is convenient to proceed from saturated solution at pressure that is larger than pressure $\Pi $ that is of interest to us. Let us identify value $n_{\infty } $ obtained during the experiment with value $n_{0} $, and then instantly reduce the pressure in the solution to $\Pi $. Due to low compressibility of the liquid solvent, the solution will become just what we are interested in -- supersaturated solution with the given concentration $n_{0} $ of the dissolved gas.

\section{Dependence of the bubble growth rate on gas solubility and solution supersaturation}
\label{GrowthRate}

In accordance with section \ref{SelfSimilarity}, self-similar Eq. \eqref{EQ_2_11_} defines parameter $b$ as an unambiguous function $b=b(a)$ of parameter $a$ throughout all the area stipulated by Eq. \eqref{EQ_2_13_}. Function $b=b(a)$ does not depend on the type of the solvent and on the type of the dissolved gas. Having multiplied function $b=b(a)$ by $2D$, using self-similar Eq. \eqref{EQ_2_7_}, we have the following equality
\begin{equation} \label{EQ_3_1_} 
dR^{2} /dt=2Db(a)\; \; \; \; \; (t{\rm \geqslant }t_{0} ) 
\end{equation} 
for the constant in time derivative $dR^{2} /dt$, which characterizes the rate of bubble growth (the rate of the growth of its surface). Function $b=b(a)$, which was obtained via numerical solution of the self-similar Eq. \eqref{EQ_2_11_}, is just what is presented in Fig. 1. It can be seen that with the increase of parameter $a$ in the area stipulated by Eq. \eqref{EQ_2_13_} this function is growing faster than parameter $a$ is.

The physical meaning of parameter $a$ is revealed in Eq. \eqref{EQ_2_14_}. Exploiting this in Eq. \eqref{EQ_3_1_}, we have
\begin{equation} \label{EQ_3_2_} 
dR^{2} /dt=2Db(s\zeta )\; \; \; \; \; (t{\rm \geqslant }t_{0} ). 
\end{equation} 
Eq. \eqref{EQ_3_2_} shows that derivative $dR^{2} /dt$ is proportional to diffusion coefficient $D$ of the molecules of gas in the solution and depends on the solubility of the gas $s$ and solution supersaturation $\zeta $ only via the product of $s\zeta $. Curve 1 in Fig. 2 represents data concerning the dependence of dimensionless value $D^{-1} dR^{2} /dt$ on $s\zeta $ obtained by means of Eq. \eqref{EQ_3_2_} and the numerical solution of the self-similar Eq. \eqref{EQ_2_11_}. It is clear that with the increase of $s\zeta $ from $s\zeta =0,01$ to $s\zeta =20$, value $D^{-1} dR^{2} /dt$ is growing at an increasing rate from $D^{-1} dR^{2} /dt=2\cdot 10^{-2} $ to $D^{-1} dR^{2} /dt=1,5\cdot 10^{3} $. The revealed effect of the significant increase of the rate of bubble growth simultaneous with the increase in the product of gas solubility $s$ and solution supersaturation $\zeta $ is nonsteady by its virtue; and this will be shown in section \ref{Steady}.

\begin{figure}
\includegraphics[width=180mm]{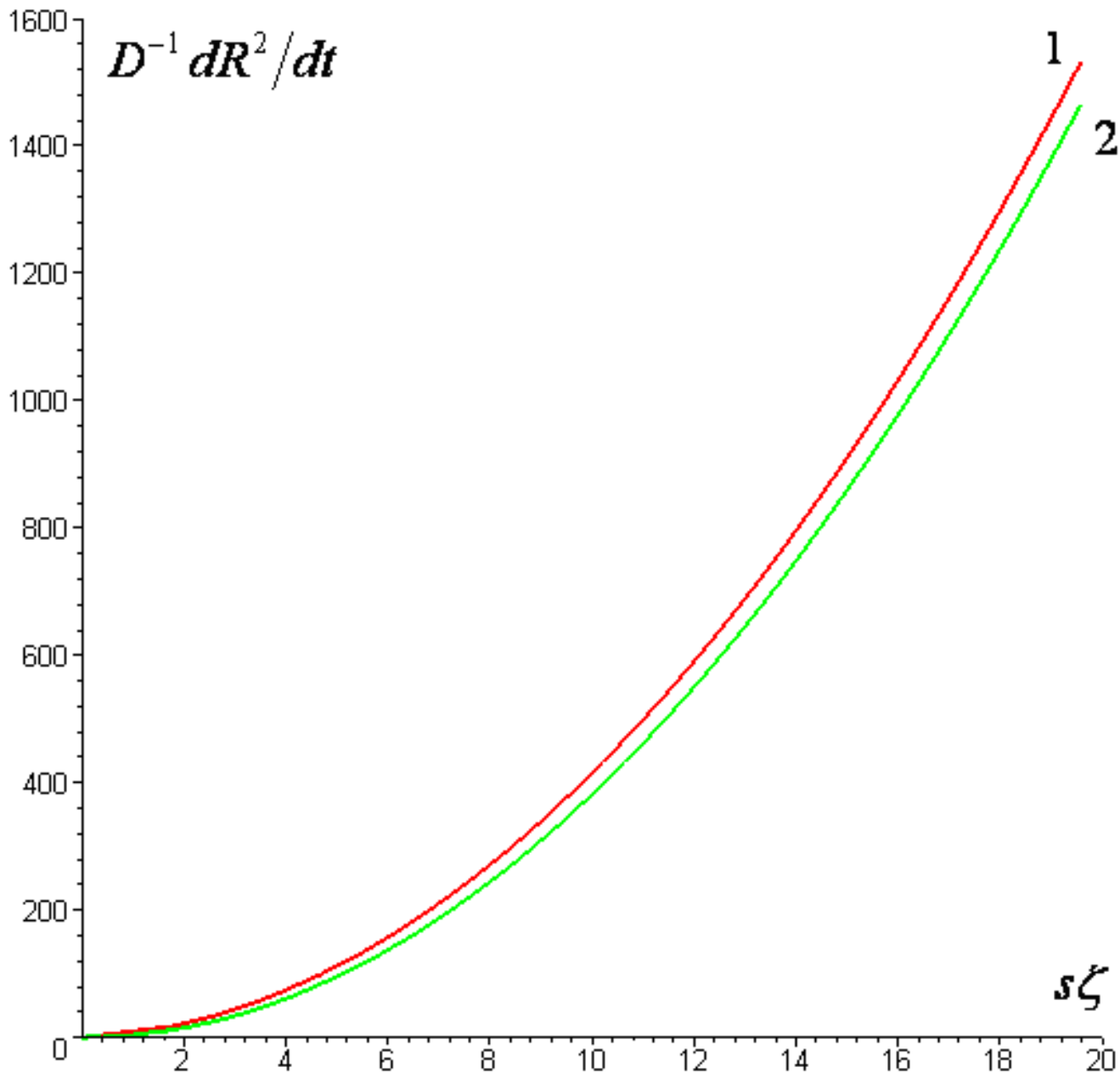}
\caption{Dependence of dimensionless value $D^{-1} dR^{2} /dt$ on the product $s\zeta $. }
\label{Fig2}
\end{figure}

When $a{\rm \gtrsim }10$, the solution of the self-similar Eq. \eqref{EQ_2_11_} can be found analytically. Let us rewrite this equation as
\begin{equation} \label{EQ_3_3_} 
\frac{a}{b} =\int _{1}^{\infty } \frac{dx}{x^{2} } \exp \left(-\frac{bx^{2} }{2} -\frac{b}{x} +\frac{3b}{2} \right). 
\end{equation} 

Expanding the exponent from Eq. \eqref{EQ_3_3_} into a Taylor series in powers of $x-1$ in the vicinity of $x=1$, noting that at $b{\rm \gg }1$ the exponent ``cuts'' the integral in Eq. \eqref{EQ_3_3_} already before $1/x^{2} $ manages to practically deviate from unit, using saddle-points technique, we have
\begin{equation} \label{EQ_3_4_} 
\int _{1}^{\infty } \frac{dx}{x^{2} } \exp \left(-\frac{bx^{2} }{2} -\frac{b}{x} +\frac{3b}{2} \right)=\left(\frac{\pi }{6} \right)^{1/2} b^{-1/2} \; \; \; \; \; (b{\rm \gtrsim }2\cdot 10^{2} ). 
\end{equation} 
Substituting Eq. \eqref{EQ_3_4_} in Eq. \eqref{EQ_3_3_}, we come up with an analytical solution of the self-similar Eq. \eqref{EQ_2_11_}: 
\begin{equation} \label{EQ_3_5_} 
b(a)=\frac{6}{\pi } a^{2} \; \; \; \; \; (a{\rm \gtrsim }10). 
\end{equation} 
Eq. \eqref{EQ_3_5_} at the limit of its applicability $a{\rm \simeq }10$ concords well with the graph given in Fig. 1. Thus analytical Eq. \eqref{EQ_3_5_} conjointly with Fig. 1 provides comprehensive data on function $b=b(a)$ throughout all the area stipulated by Eq. \eqref{EQ_2_13_}. Curve 2 in Fig. 2 represents data on the dependence of dimensionless value $D^{-1} dR^{2} /dt$ on $s\zeta $ obtained by means of Eq. \eqref{EQ_3_2_} and the analytical Eq. \eqref{EQ_3_5_}. 

Exploiting Eq. \eqref{EQ_3_5_} in Eq. \eqref{EQ_3_2_}, we have 
\begin{equation} \label{EQ_3_6_} 
\frac{dR^{2} }{dt} =\frac{12D}{\pi } (s\zeta )^{2} \; \; \; \; \; (s\zeta {\rm \gtrsim }10,\; t{\rm \geqslant }t_{0} ). 
\end{equation} 
Analytical formula \eqref{EQ_3_6_} shows that the derivative $dR^{2} /dt$ is growing rather quickly with the growth of $s\zeta $ at $s\zeta {\rm \gtrsim }10$.

Let us assess time $t_{0} $ in the condition $t{\rm \geqslant }t_{0} $ of the applicability of Eqs. \eqref{EQ_3_1_}, \eqref{EQ_3_2_} and \eqref{EQ_3_6_}. Let us rewrite $t_{0} $ in accordance with Eqs. \eqref{EQ_2_17_} and \eqref{EQ_2_7_} as
\begin{equation} \label{EQ_3_7_} 
t_{0} {\rm \simeq }R_{0}^{2} /(dR^{2} /dt). 
\end{equation} 
Eq. \eqref{EQ_3_7_} shows that value represents the rate of relative increase of the area of the bubble surface at time $t_{0} $.

For radius $R_{0} $ introduced by means of Eq. \eqref{EQ_1_6_}, let us write
\begin{equation} \label{EQ_3_8_} 
R_{0} {\rm \sim }40\frac{\sigma }{\Pi } , 
\end{equation} 
where it is considered that the largest of the two values given in braces in Eq. \eqref{EQ_1_6_} is, as a rule, value $2\sigma /\Pi $. In accordance with Eq. \eqref{EQ_3_8_}, relative deviation of pressure in the bubble of radius $R{\rm \geqslant }R_{0} $ from pressure $\Pi $ of the solution does not exceed $5\% $. Let us assume that $\Pi {\rm \simeq }10^{6} \; \frac{{\rm dyne}}{{\rm cm}^{{\rm 2}} } {\rm \simeq }1\; {\rm atm}$ and $\sigma {\rm \simeq }75\; \frac{{\rm dyne}}{{\rm cm}} $ (as in case of water). Then, from Eq. \eqref{EQ_3_8_} we have
\begin{equation} \label{EQ_3_9_} 
R_{0} {\rm \sim }3\cdot 10^{-3} \; {\rm cm}. 
\end{equation} 

Let us find the derivative $dR^{2} /dt$ in Eq. \eqref{EQ_3_7_} by means of Fig. 2, using the value that is typical for water: $D{\rm \simeq }1,6\cdot 10^{-5} \; \frac{{\rm cm}^{{\rm 2}} }{{\rm s}} $ \cite{Handbook}. Further we shall obtain $R_{0}^{2} $ in Eq. \eqref{EQ_3_7_} using the evaluation from Eq. \eqref{EQ_3_9_}. The data on the dependence of the evaluation of time $t_{0} $ on $s\zeta $ obtained in the described way are presented in Fig. 3.
\begin{figure}
\includegraphics[width=180mm]{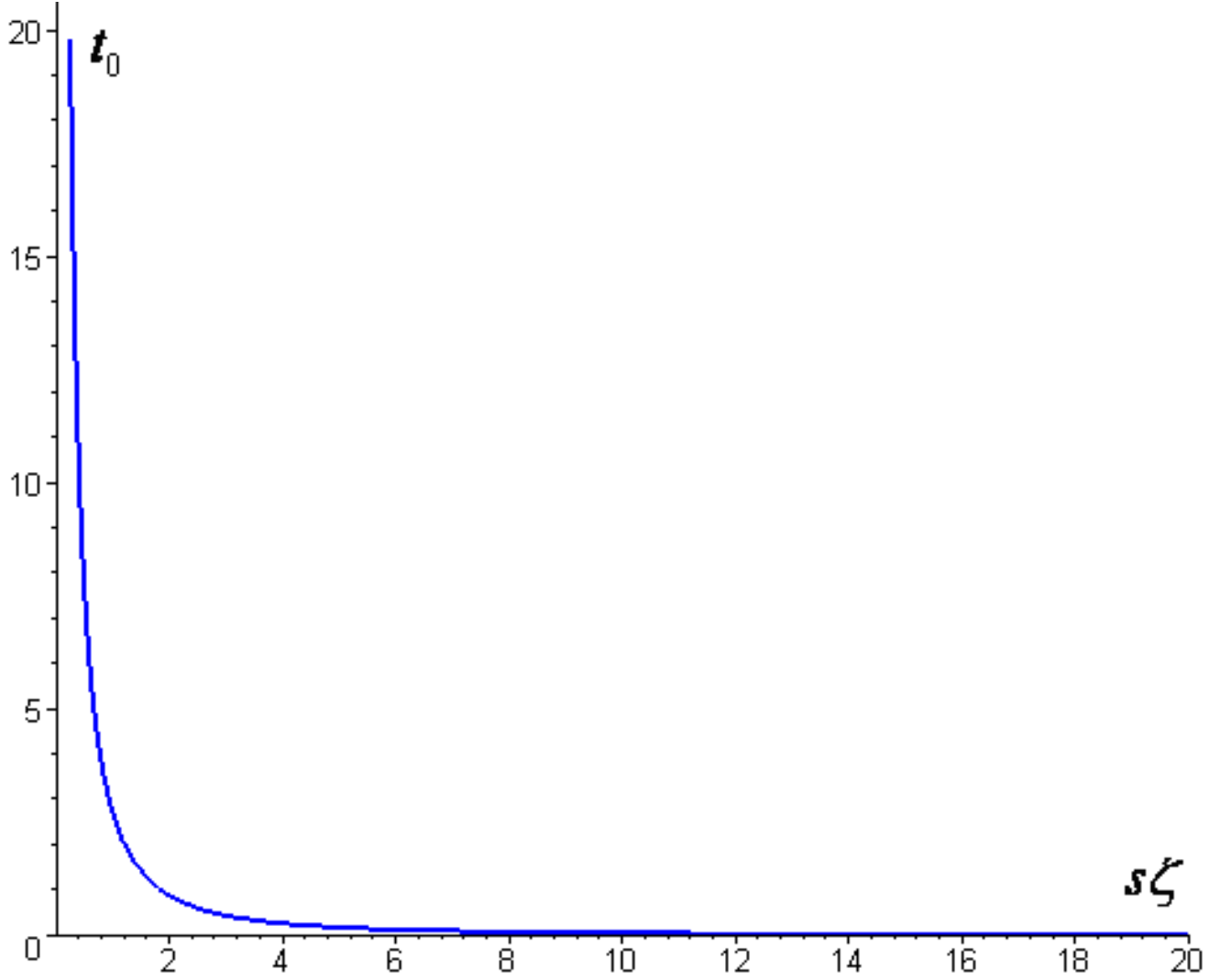}
\caption{Dependence of the estimation for the time $t_0 ~(s)$ on the product $s\zeta $. }
\label{Fig3}
\end{figure}

Using the analytical Eq. \eqref{EQ_3_6_} in Eq. \eqref{EQ_3_7_} at the same value of diffusion coefficient $D$ and the same evaluation from Eq. \eqref{EQ_3_9_}, we have 
\begin{equation} \label{EQ_3_10_} 
t_{0} {\rm \sim }\frac{1,5\cdot 10^{-1} }{(s\zeta )^{2} } \; {\rm s}\; \; \; \; \; (s\zeta {\rm \gtrsim }10). 
\end{equation} 
Analytical Eq. \eqref{EQ_3_10_} and Fig. 3 provide exhaustive data on the evaluation of time $t_{0} $ throughout all the area of $s\zeta >0$.

Let us notice that the Eq. \eqref{EQ_2_18_} can be rewritten using Eq. \eqref{EQ_2_17_} as
\begin{equation} \label{EQ_3_11_} 
R^{2} (t){\rm \simeq }R_{0}^{2} \; t/t_{0} \; \; \; \; \; (t{\rm \geqslant }t_{0} ),
\end{equation} 
where values $R_{0} $ and $t_{0} $ evaluated previously are present.

\section{Steady theory case}
\label{Steady}

Let us study a case when
\begin{equation} \label{EQ_4_1_} 
\left(s\zeta \right)^{1/2} {\rm \ll }1. 
\end{equation} 
Thus, from Eq. \eqref{EQ_2_14_} it follows that
\begin{equation} \label{EQ_4_2_} 
a^{1/2} {\rm \ll }1. 
\end{equation} 
As we shall see further, inequality in Eq. \eqref{EQ_4_2_} will help us to solve a self-similar Eq. \eqref{EQ_2_11_} analytically. The same inequality will also stipulate the range of applicability of the steady theory.

Let us begin by solving Eq. \eqref{EQ_2_11_}. Differentiating by $b$, one can make sure that the following asymptotic equation is observed:
\begin{equation} \label{EQ_4_3_} 
\int _{1}^{\infty } \frac{dx}{x^{2} } {\rm e}^{-{bx^{2} \mathord{\left/ {\vphantom {bx^{2}  2-{b\mathord{\left/ {\vphantom {b x}} \right. \kern-\nulldelimiterspace} x} }} \right. \kern-\nulldelimiterspace} 2-{b\mathord{\left/ {\vphantom {b x}} \right. \kern-\nulldelimiterspace} x} } } =1-\left(\frac{\pi }{2} \right)^{1/2} b^{1/2} +...\; \; \; \; \; (b^{1/2} {\rm \ll }1), 
\end{equation} 
where the neglected components in terms of their value are even smaller than $b$. Using Eq. \eqref{EQ_4_3_} in a self-similar Eq. \eqref{EQ_2_11_}, considering only its main component, we shall have $a=b$, which, by virtue of Eq. \eqref{EQ_4_2_}, confirms the condition $b^{1/2} {\rm \ll }1$ in Eq. \eqref{EQ_4_3_}. As a result, the solution of a self-similar Eq. \eqref{EQ_2_11_}, in case when Eq. \eqref{EQ_4_2_} is observed, in the principal order by $a^{1/2} $ looks as follows:
\begin{equation} \label{EQ_4_4_} 
b=a\; \; \; \; \; (a^{1/2} {\rm \ll }1). 
\end{equation} 

Now let us define the range of applicability of the steady theory. Let us substitute Eq. \eqref{EQ_4_4_} in the self-similar Eq. \eqref{EQ_2_10_}. Thus we have 
\begin{equation} \label{EQ_4_5_} 
n(\rho )=n_{\infty } +n_{g} a\int _{1}^{\rho } \frac{dx}{x^{2} } \exp \left(-\frac{ax^{2} }{2} -\frac{a}{x} +\frac{3a}{2} \right)\; \; \; \; \; (a^{1/2} {\rm \ll }1). 
\end{equation} 
Let us prove that the concentration profile of the dissolved gas in Eq. \eqref{EQ_4_5_}, which was obtained at the limitation from above on $a$ expressed in Eq. \eqref{EQ_4_2_}, is steady in the area of
\begin{equation} \label{EQ_4_6_} 
1{\rm \leqslant }\rho {\rm \lesssim }2^{1/2} /4a^{1/2} \; \; \; \; \; (a^{1/2} {\rm \ll }1), 
\end{equation} 
that exists simultaneously with limitation from Eq. \eqref{EQ_4_2_}. In the area stipulated by Eq. \eqref{EQ_4_6_} the absolute value in the exponent in Eq. \eqref{EQ_4_5_} is rather small (smaller than $1/16$), and the exponent itself, consequently, is rather close to $1$. Taking this fact into consideration and also developing $a$ in the multiplier in front of the integral in Eq. \eqref{EQ_4_5_}, using definition from Eq. \eqref{EQ_2_12_}, in the area stipulated by Eq. \eqref{EQ_4_6_} we shall with high precision deduce Eq. \eqref{EQ_4_5_} to
\begin{equation} \label{EQ_4_7_} 
n(\rho )=n_{0} -\frac{n_{0} -n_{\infty } }{\rho } \; \; \; \; \; (a^{1/2} {\rm \ll }1,\; \; \; 1{\rm \leqslant }\rho {\rm \lesssim }2^{1/2} /4a^{1/2} ). 
\end{equation} 
Eq. \eqref{EQ_4_7_} corresponds to the steady concentration profile of the dissolved gas. We also provide limitations on $a$ and on $\rho $; when those limitations are observed Eq. \eqref{EQ_4_7_} stems from a general self-similar Eq. \eqref{EQ_2_10_} for the nonsteady concentration profile of the dissolved gas.

In order to find the rate of bubble growth, as it can be seen from Eq. \eqref{EQ_2_5_}, it is enough to know the derivative $dn(\rho )/d\rho |_{\rho =1} $, i.e. it is enough to know the concentration profile of the dissolved gas only in the infinitely narrow vicinity of the bubble. But even then one can obtain this data within the framework of the steady approximation of Eq. \eqref{EQ_4_7_} only when the limitation Eq. \eqref{EQ_4_2_} is observed. Indeed, already when $a^{1/2} {\rm \gtrsim }2^{1/2} /4$ (when the limitation of Eq. (4.2) is no longer observed) the area stipulated by Eq. \eqref{EQ_4_6_} does not exist at all; and, consequently, there is no such thing as an infinitely narrow vicinity of the bubble, where the concentration profile of the dissolved gas could be steady.

We see that the limitation in Eq. \eqref{EQ_4_2_} is necessary for works \cite{Kuni_Ogenko_1, Trofimov_Melikhov_Kuni, Kuni_Zhuvikina_2002, Kuni_Zhuvikina_Grinin, Slezov_1, Slezov_2} based on the steady theory and devoted to the study of bubble growth. At the same time, it becomes clear that the rapid increase in the rate of bubble growth, which, as it was revealed in section \ref{GrowthRate}, occurs at the increase of the product $s\zeta $, and, consequently, at the increase of value $a$ (when Eq. (4.2) is not observed) , is of a significantly nonsteady nature.

Finally, let us apply the analytical solution from Eq. \eqref{EQ_4_4_} of the self-similar Eq. \eqref{EQ_2_11_} to the problem of finding the rate of bubble growth. Substituting Eq. \eqref{EQ_4_4_} in the self-similar Eq. \eqref{EQ_2_7_}, we have 
\begin{equation} \label{EQ_4_8_} 
dR^{2} /dt=2Da\; \; \; \; \; (a^{1/2} {\rm \ll }1) 
\end{equation} 
(for the sake of simplicity we do not point out the inequality $t{\rm \geqslant }t_{0} $ which is the condition for the observation of Eq. \eqref{EQ_2_7_}). Exploiting Eq. \eqref{EQ_2_14_} in Eq. \eqref{EQ_4_8_}, we have
\begin{equation} \label{EQ_4_9_} 
dR^{2} /dt=2Ds\zeta \; \; \; \; \; ((s\zeta )^{1/2} {\rm \ll }1). 
\end{equation} 

Using the definition from Eq. \eqref{EQ_1_3_}, Eq. \eqref{EQ_4_9_} can be rewritten as
\begin{equation} \label{EQ_4_10_} 
\frac{dR}{dt} =\frac{kTn_{\infty } D}{R\Pi } \zeta . 
\end{equation} 
Under the condition that $P_{\beta } {\rm \ll }\Pi $ specified in the beginning of section \ref{GeneralIdeas}, Eq. \eqref{EQ_4_10_} coincides with Eq. (2.7) in \cite{Kuni_Zhuvikina_Grinin} obtained at low gas solubility.

\section{Range of applicability of the theory}
\label{Applicability}

Let us first find the condition on the smallness of the deviation of the bubble temperature from the temperature of the solution. The transition of a molecule of gas from the solution to the bubble, as a rule, occurs with absorption of heat. As a result, the temperature of the bubble, which we will denote as $T_{g} $, becomes smaller than temperature $T$ of the solution (now we perceive $T$ as the temperature of the solution at the infinite distance from the bubble). In order to find the deviation $T_{g} -T$, one preliminary has to find the nonsteady temperature profile $T(r,t)$ around the bubble by solving the heat conductivity equation in the liquid solvent moving outside the bubble with the boundary condition $T(r,t)|_{r=\infty } =T$ and at the intensity of absorption of heat by the bubble stipulated by the rate of its growth. This can be achieved, just as in section \ref{SelfSimilarity}, by means of a self-similar method based on the substitution $T(r,t)=T(\rho )$, where a self-similar variable $\rho $ is given by the previously used equality \eqref{EQ_2_1_}. (Study of nonsteady heat conductivity problem is important for the growth of a liquid droplet in supersaturated vapor \cite{Grinin_Gor_Kuni}). Using this method and obtaining $T(\rho )$, we will have the following equality for the unknown deviation $T_{g} -T$, where $T_{g} $ is defined as $T_{g} \equiv T(\rho )|_{\rho =1} $:
\begin{equation} \label{EQ_5_1_} 
T_{g} -T=-\frac{qDn_{g} }{\kappa } b\int _{1}^{\infty } \frac{dx}{x^{2} } \exp \left(-\frac{\varepsilon bx^{2} }{2} -\frac{\varepsilon b}{x} +\frac{3\varepsilon b}{2} \right). 
\end{equation} 
Here $q$ is the heat of dissolution per one molecule of gas (as a rule, $q>0$), $\kappa $ is the coefficient of heat conductivity of the liquid solvent in the diluted solution, 
\begin{equation} \label{EQ_5_2_} 
\varepsilon \equiv D/\chi , 
\end{equation} 
$\chi $ is the coefficient of thermal diffusivity of the liquid solvent in the diluted solution. For the rate of bubble growth in Eq. \eqref{EQ_5_1_} we have used Eq. \eqref{EQ_2_7_}, which is enough to assert the smallness of heat effect, (but not enough to study it).

Replacing value $b$ with value $\varepsilon b$ in Eq. \eqref{EQ_3_4_}, we have
\begin{equation} \label{EQ_5_3_} 
\int _{1}^{\infty } \frac{dx}{x^{2} } \exp \left(-\frac{\varepsilon bx^{2} }{2} -\frac{\varepsilon b}{x} +\frac{3\varepsilon b}{2} \right)=\left(\frac{\pi }{6} \right)^{1/2} (\varepsilon b)^{-1/2} \; \; \; \; \; (\varepsilon b{\rm \gtrsim }2\cdot 10^{2} ). 
\end{equation} 
From Eqs. \eqref{EQ_5_1_} and \eqref{EQ_5_3_} we have 
\begin{equation} \label{EQ_5_4_} 
T_{g} -T=-\frac{qDn_{g} }{\varepsilon ^{1/2} \kappa } \left(\frac{\pi }{6} \right)^{1/2} b^{1/2} \; \; \; \; \; (b{\rm \gtrsim }2\cdot 10^{2} /\varepsilon ). 
\end{equation} 

We shall assume values characteristic for water in ordinary conditions \cite{Handbook}:
\begin{equation} \label{EQ_5_5_} 
D{\rm \simeq }1,6\cdot 10^{-5} \; \frac{{\rm cm}^{{\rm 2}} }{{\rm s}} ,\; \; \kappa {\rm \simeq }6\cdot 10^{4} \; \frac{{\rm erg}}{{\rm cm\times s\times K}} ,\; \; \chi {\rm \simeq }1,4\cdot 10^{-3} \; \frac{{\rm cm}^{{\rm 2}} }{{\rm s}} . 
\end{equation} 
Then, for value $\varepsilon $ introduced in Eq. \eqref{EQ_5_2_} we have 
\begin{equation} \label{EQ_5_6_} 
\varepsilon {\rm \simeq }10^{-2} . 
\end{equation} 

As can be seen from Eqs. \eqref{EQ_3_5_} and \eqref{EQ_5_6_}, limitation $b{\rm \gtrsim }2\cdot 10^{2} /\varepsilon $ in Eq. \eqref{EQ_5_4_} is equipotent to limitation $a{\rm \gtrsim }10^{2} $ which makes Eq. \eqref{EQ_3_5_} correct. Then, in accordance with Eqs. \eqref{EQ_5_4_}, \eqref{EQ_5_6_} and \eqref{EQ_3_5_}, we have
\begin{equation} \label{EQ_5_7_} 
T_{g} -T{\rm \simeq }-\frac{10qDn_{g} }{\kappa } a\; \; \; \; \; (a{\rm \gtrsim }10^{2} ). 
\end{equation} 
Using Eq. \eqref{EQ_5_7_} we have 
\begin{equation} \label{EQ_5_8_} 
\frac{|T_{g} -T|}{T} {\rm \simeq }\frac{10|q|Dn_{g} }{\kappa T} a\; \; \; \; \; (a{\rm \gtrsim }10^{2} ). 
\end{equation} 

Let us accept that
\begin{equation} \label{EQ_5_9_} 
\Pi {\rm \simeq }10^{6} \; \frac{{\rm dyne}}{{\rm cm}^{{\rm 2}} } ,\; \; T_{g} {\rm \simeq }300\; {\rm K},\; \; |q|/kT{\rm \simeq }10. 
\end{equation} 
We will use Eqs. \eqref{EQ_5_5_}, \eqref{EQ_5_9_} in Eq. \eqref{EQ_5_8_}, and then expand $n_{g} $ by means of Eq. \eqref{EQ_1_4_}. Thus we will have 
\begin{equation} \label{EQ_5_10_} 
|T_{g} -T|/T{\rm \simeq }10^{-4} a\; \; \; \; \; (a{\rm \gtrsim }10^{2} ). 
\end{equation} 
From Eq. \eqref{EQ_5_10_} it follows that 
\begin{equation} \label{EQ_5_11_} 
|T_{g} -T|/T{\rm \lesssim }10^{-1} \; \, \, \, \, {\rm when}\; \, \, \, \, \, a{\rm \lesssim }10^{3} , 
\end{equation} 
where we do not refer to limitation $a{\rm \gtrsim }10^{2} $ in Eq. \eqref{EQ_5_10_}, as in accordance with physical meaning the value in the left part of Eq. \eqref{EQ_5_11_} decreases with the decrease of value $a$. Inequality in Eq. \eqref{EQ_5_11_} reveals the smallness of the relative deviation of the bubble temperature from the temperature of the solution even in the case of very large values of parameter $a=s\zeta {\rm \lesssim }10^{3} $, where Eq.\eqref{EQ_2_14_} is observed.

Now we will obtain the condition on the equalization of the temperature of the bubble. For time $t_{g} $ which is characteristic for this equalization we have the following estimate: $t_{g} {\rm \sim }R^{2} /\chi _{g} $, where $\chi _{g} $ is the coefficient of thermal diffusivity of gas in the bubble. Let us now introduce characteristic time $t_{R} =R/(dR/dt)$ for the changing of the bubble radius. The fact that the temperature of the bubble manages to establish in the process of the bubble growth is equivalent to strong inequality $t_{g} {\rm \ll }t_{R} $, that is
\begin{equation} \label{EQ_5_12_} 
R/\chi _{g} {\rm \ll }1/(dR/dt). 
\end{equation} 
By means of self-similar Eq. \eqref{EQ_2_6_} we shall rewrite Eq. \eqref{EQ_5_12_} as 
\begin{equation} \label{EQ_5_13_} 
Db/\chi _{g} {\rm \ll }1. 
\end{equation} 
Using Eq. \eqref{EQ_3_5_}, we have 
\begin{equation} \label{EQ_5_14_} 
\frac{Db}{\chi _{g} } =\frac{6}{\pi } \frac{D}{\chi _{g} } a^{2} \; \; \; \; \; \; (a{\rm \gtrsim }10). 
\end{equation} 
From Eq. \eqref{EQ_5_14_}, taking Eq. \eqref{EQ_5_5_} and the fact that $\chi _{g} {\rm \sim }10^{-1} \; \frac{{\rm cm}^{{\rm 2}} }{{\rm s}} $ into consideration, it follows that 
\begin{equation} \label{EQ_5_15_} 
Db/\chi _{g} {\rm \lesssim }10^{-1} \, \, \, \; {\rm when}\, \, \, \, a{\rm \lesssim }1,8\cdot 10, 
\end{equation} 
where we do not refer to limitation $a{\rm \gtrsim }10$ in Eq. \eqref{EQ_5_14_}, as value $b$ decreases with the decrease of value $a$. Inequality in Eq. \eqref{EQ_5_15_} reveals the smallness of value $Db/\chi _{g} $in the condition from Eq. \eqref{EQ_5_13_} at all values of $s\zeta {\rm \lesssim }1,8\cdot 10$, where Eq.\eqref{EQ_2_14_} is observed.

Thus, if we consider conditions on the isothermal character of the bubble growth in Eqs. \eqref{EQ_5_11_} and \eqref{EQ_5_15_}, the latter condition turns out to be stronger, that is 
\begin{equation} \label{EQ_5_16_} 
s\zeta {\rm \lesssim }1,8\cdot 10. 
\end{equation} 
Still it leaves broad possibilities for the application of the isothermal theory. Indeed, the condition from Eq. \eqref{EQ_5_16_} in accordance with Eq. \eqref{EQ_3_6_} admits even that
\begin{equation} \label{EQ_5_17_} 
0<D^{-1} dR^{2} /dt{\rm \lesssim }1,2\cdot 10^{3} \; \; \; \; \; \; (s\zeta {\rm \lesssim }1,8\cdot 10,\; \; t{\rm \geqslant }t_{0} ), 
\end{equation} 
where the limitation $s\zeta {\rm \gtrsim }10$ in Eq. \eqref{EQ_3_6_} is not mentioned as $D^{-1} dR^{2} /dt$ decreases with the decrease of $s\zeta $. As for velocity $1/t_{0} $ of the relative increase of the area of the bubble surface at time $t_{0} $, the condition in Eq. \eqref{EQ_5_16_} in accordance with Eq. \eqref{EQ_3_10_} admits even that
\begin{equation} \label{EQ_5_18_} 
0<1/t_{0} {\rm \lesssim }2,3\cdot 10^{3} \; {\rm s}^{{\rm -1}} \; \; \; \; \; \; (s\zeta {\rm \lesssim }1,8\cdot 10), 
\end{equation} 
where limitation $s\zeta {\rm \gtrsim }10$ in Eq. \eqref{EQ_3_10_} is not mentioned as $1/t_{0} $ decreases with the decrease of $s\zeta $. The ranges of alteration of values $D^{-1} dR^{2} /dt$ è $1/t_{0} $ â \eqref{EQ_5_17_} è \eqref{EQ_5_18_} comprise three orders of magnitude. Therefore, the possibilities of the isothermal theory are rather broad.

Let us prove that the observation of the condition from Eq. \eqref{EQ_5_16_} causes the observation of the condition on the diluteness of the solvent. In accordance with Eqs. \eqref{EQ_1_4_}, \eqref{EQ_1_5_} and \eqref{EQ_1_18_}, we have 
\begin{equation} \label{EQ_5_19_} 
n_{0} =(s\zeta +s)\Pi /kT. 
\end{equation} 
As $s{\rm \ll }10$ is correct in all cases, then from Eq. \eqref{EQ_5_19_} in accordance with Eq. \eqref{EQ_5_9_} it follows that 
\begin{equation} \label{EQ_5_20_} 
n_{0} {\rm \lesssim }4,3\cdot 10^{20} \; {\rm cm}^{-3} \; \; \; \; \; \; (s\zeta {\rm \lesssim }1,8\cdot 10). 
\end{equation} 
The estimation from Eq. \eqref{EQ_5_20_} shows that, when the condition in Eq. \eqref{EQ_5_16_} is observed, the concentration $n_{0} $ of the dissolved gas is approximately two orders less that the concentration $n_{l} $ of the liquid solvent ($n_{l} {\rm \simeq }3\cdot 10^{22} \; {\rm cm}^{-3} $ for the case of water). This substantiates the diluteness of the solvent.

Let us finally find the condition which is responsible for the observation of the mechanical equilibrium of the bubble and the solvent which was presupposed in the paper. We will use $t_{h} $ to denote the time of mechanical relaxation of the bubble under the influence of the hydrodynamic interaction between the bubble and the solvent. As we consider large bubbles, i. e. cases when, in accordance with the estimation in Eq. \eqref{EQ_3_9_}, $R{\rm \gtrsim }3\cdot 10^{-3} \; {\rm cm}$, the following inequality is observed
\begin{equation} \label{EQ_5_21_} 
3\Pi \rho _{p} R^{2} /4\eta ^{2} >1, 
\end{equation} 
where $\rho _{p} $ and $\eta $ are mass density and  the viscosity of the solvent accordingly. Indeed, this fact is obvious from $\Pi {\rm \simeq }10^{6} \; \frac{{\rm dyne}}{{\rm cm}^{{\rm 2}} } $, $\rho _{p} =1\; \frac{{\rm g}}{{\rm cm}^{{\rm 3}} } $ and $\eta =10^{-2} \; \frac{{\rm g}}{{\rm cm\times s}} $. As it has been shown in \cite{Trofimov_Melikhov_Kuni}, when inequality in Eq. \eqref{EQ_5_21_} is observed, the mechanical relaxation of the bubble is of oscillation and attenuation character, while time $t_{h} $ of this relaxation is given by the following equality: 
\begin{equation} \label{EQ_5_22_} 
t_{h} =\rho _{p} R^{2} /2\eta  
\end{equation} 
(Eq. (23) in \cite{Trofimov_Melikhov_Kuni}). Let us introduce characteristic time $t_{R} =R/(dR/dt)$ of the alteration of the radius of the bubble. The condition on the observation of the mechanical equilibrium of the bubble and the solvent is 
\begin{equation} \label{EQ_5_23_} 
t_{h} /t_{R} {\rm \ll }1. 
\end{equation} 
Obtaining $t_{R} $ by means of Eq. \eqref{EQ_3_6_} and taking Eq, \eqref{EQ_5_22_} into account, we have 
\begin{equation} \label{EQ_5_24_} 
\frac{t_{h} }{t_{R} } =\frac{3\rho _{p} D}{\pi \eta } (s\zeta )^{2} \; \; \; \; \; \; (s\zeta {\rm \gtrsim }10). 
\end{equation} 
From Eq. \eqref{EQ_5_24_}, when$D{\rm \simeq }1,6\cdot 10^{-5} \; \frac{{\rm cm}^{{\rm 2}} }{{\rm s}} $, $\rho _{p} =1\; \frac{{\rm g}}{{\rm cm}^{{\rm 3}} } $ and $\eta =10^{-2} \; \frac{{\rm g}}{{\rm cm\times s}} $ 
\begin{equation} \label{EQ_5_25_} 
t_{h} /t_{R} <5\cdot 10^{-1} \; \; \; \; \; \; (s\zeta {\rm \lesssim }1,8\cdot 10), 
\end{equation} 
where we do not point at limitation $s\zeta {\rm \gtrsim }10$ in Eq. \eqref{EQ_5_24_}, as in accordance with physical meaning value $t_{h} /t_{R} $ decreases with the decrease of value $s\zeta $. In accordance with Eq. \eqref{EQ_5_25_}, the inequality in Eq. \eqref{EQ_5_23_} is observed in the vicinity of the boundary value even under the condition of Eq. \eqref{EQ_5_16_} which, as it has been mentioned earlier provided extensive possibilities for the application of the theory. We will impose this condition on $s\zeta $ in Eq. \eqref{EQ_5_25_} in order to reveal it that the mechanical equilibrium between the bubble and the solvent and the equalization of temperature inside the bubble cease being observed at approximately the same values of $s\zeta $.

In conclusion we will note that the smallness of gas solubility was not exploited in the present paper.

\begin{acknowledgments}
The authors are grateful to Professor L. Ts. Adzhemyan and Professor A. K. Shchekin for discussions and helpful remarks they made. The authors are also indebted to Professor A. E. Kuchma for his critical advises on the manuscript. The research has been carried out with the financial support of the Russian Analytical Program ``The Development of Scientific Potential of Higher Education'' (2006 -- 2008), project RNP.2.1.1.1812. Fundamental Problems of Physics and Chemistry of Ultradisperse Systems and Interfaces.\end{acknowledgments}

\newpage
\bibliography{Manuscript}

\end{document}